\documentclass[11pt]{article}
\usepackage{epsfig}
\csname @addtoreset\endcsname{equation}{section}
\def\be{\begin{equation}}
\def\ee{\end{equation}}
\def\ba{\begin{eqnarray}}
\def\ea{\end{eqnarray}}
\def\no{\nonumber \\}

\begin{document}

{\Large 
\begin{center}
{\bf{Wilson Loop and Dimensional Reduction in \\ Non-Commutative Gauge Theories}} 
\end{center}}
\vskip 2cm
\begin{center}
Sunggeun Lee\footnote{sglee@hepth.hanyang.ac.kr} and Sang-Jin Sin
\footnote{sjs@hepth.hanyang.ac.kr}\\ \vskip 1cm 
{\it{Department of Physics, Hanyang University, Seoul, Korea}}
\end{center}
\vspace{0.5in}
\begin{center}
\large{\bf{Abstract}}
\end{center}
Using the AdS/CFT correspondence we study UV behavior of Wilson loops 
in various noncommutative gauge theories. We get an area law in most 
cases and try to identify its origin. In D3 case, we may identify the the 
origin as the D1 dominance over the D3: as we go to the boundary of the AdS space,
the effect of the flux of the D3 charge is highly suppressed, 
while the flux due to the D1 charge is enhenced. So near the boundary the theory is more like 
a theory on D1 brane than that on D3 brane. 
This phenomena is closely related to the dimensional reduction 
due to the strong magnetic field  in the charged particle in the magnetic field. 
The linear potential is not due to the 
confinement by IR effect but is the analogue of Coulomb's potential in  1+1  
dimension. 

\vskip 1.0cm
\renewcommand{\theequation}{\thesection.\arabic{equation}}
\pagenumbering{arabic}
\newpage
%%%%%%%%%%%%%%%%%%%%%%%%%%%%%%%%%%%%%%%%%%%%%%%%%%%%%%%%%%%%%%%%%
\section{Introduction}
%%%%%%%%%%%%%%%%%%%%%%%%%%%%%%%%%%%%%%%%%%%%%%%%%%%%%%%%%%%%%%%%%

Recently, Maldacena\cite{mal} (See \cite{agmoo,pdiv} for a review) 
conjectured that string theory on AdS space-time is dual to 
SU(N) SYM, named AdS/CFT correspondence. If 
we turn on a NS $B$ field on the N folding $D$-brane world 
volume, the  low energy effective theory is equivalent to a 
noncommutative U(N) Super Yang-Mills(NCSYM) theory  \cite{dh,cds,aas,sj,vs,ch,sw}. 
The corresponding  dual gravity solution with nonvanishing $B$ field was constructed
 in \cite{tdual} as a bound state of branes.  

A Wilson loop can be calculated by the minimal area whose 
boundary is the given Wilson loop\cite{wloop}.
In non-vanishing $B$-field background, 
 it is observed that
the string tilts from its usual direction (orthogonal to the boundary of 
AdS) by certain angle so that the length of the string along the boundary is 
infinite\cite{mr}. Wilson 
loop that goes deep into the  near horizon (IR) was found 
to give a Coulombic potential.  In \cite{aos}, it was observed that for 
a string moving with special velocity, the   tilting angle is 
zero and the effect of the non-commutativity is merely  renormalizing 
the Coulomb potential. So far, however, it is not clear why one should 
 calculate the Wilson loop behavior at a fine tuned velocity. 
The unusual feature of the Wilson loops in the presence of the 
$B$ field are  associated with non-locality of the boundary gauge 
theory and the lack of the gauge invariance of the Wilson loop\cite{das1,das2,ghi}. The super gravity 
solution is not asymptotically $AdS_5$ space: the noncommutative directions
shrink near the boundary. So there are some skepticism whether one 
can extract any physics out of the Wilson loop in non-commutative gauge theory.

In recent paper of  Dhar and Kitazawa \cite{dk}, it is noticed
  that if we place the boundary at the finite position $u=\Lambda$,
we can find a branch that gives the Coulomb potential and the situation looks  
as a small deformation of the commutative case. The price for 
having such branch is that the string configuration is not uniquely 
determined for a given length of Wilson line unless one put the probe brane at 
the non-commutative scale, $u\sim 1/a $. If we put the probe brane at $1/a$, 
we are cutting out all the "non-commutative region" (strong $B$-field region) in the bulk. 
 Therefore it is not surprising that 
they  get the Coulomb's law for large Wilson line. 
Small Wilson line, whose Nambu-Goto string stays near the 
boundary,  can 'feel' the effect of the $B$ field.
In this case they find the area law. So they got the transition from 
Coulomb to area law as the size of the wilson line changes from 
large to small. Although interesting, 
the physics of the area law is not clear at all in this approach, 
especially because they cut out the all the strong B-field region. 

In this paper, we try to identify the mechanism of the area law. 
If the area law is a character of the 
the non-commutativity, we can expect that we should get it 
for any Wilson line which stay in the large $B$ field region. 
So we do not put the boundary at the finite $u$.
We put it at infinity as usual. 
As a consequence, the Coulomb branch, is not available to 
us. We will probe the non-commutative regime where the minimum point of the string, $u_0$, is larger
than the non-commutatve scale, $1/a$, so 
that entire Nambu-Goto string of the Wilson line is in the strong 
$B$-field region. We will find that the Wilson line follows universal 
area law. This is contrasted with the case of commutative case,  where
 temporal loop gives Coulomb's law while spatial loop 
gives an area law \cite{mal,pdiv,fiwil,sonn,wit}. In the presence $B$ field case,
we will show that we get area law for both case.

In D3 case, we may identify the the 
origin as the D1 dominance \cite{roy,cai} over the D3: as we go to the boundary of the AdS space,
the effect of the flux of the D3 charge is highly suppressed, 
while the flux due to the D1 charge is enhenced. So near the boundary, the theory is more like 
a theory on D1 brane than that on D3 brane. 
This phenomena is closely related to the dimensional reduction 
due to the strong magnetic field  in the charged particle in the magnetic field. 
Then, the linear potential is not due to the 
confinement by IR effect but is the 'analogue' of Coulomb's potential in  1+1  
dimension. 

This paper is organized as follows. In section 2, we review the 
gravity dual of non-commutative gauge theory and its scaling 
symmetries. In section 3, we calculate the Wilson loop in UV regime for various cases including 
finite temperatures, spatial as well as temporal loops, 
$D$-instanton background and other $D_p$ brane cases. We get an area law almost universally
if time is not  non-commutative. In section 4, we give a 
physical interpretation for the area law of 3+1 dimensional non-commutative gauge 
theories as D1 dominance and  dimensional reduction due to the magnetic field.
 We summarize and conclude in section 5. 
 
\section{ Gravity dual of the non-commutative gauge 
theory and its scaling symmetry}
 Let us first consider the zero temperature case of  
$D3$ brane with constant $B$ field   parallel 
to the brane. Its low energy effective world-volume theory is 
described by noncommutative Yang-Mills theory. The gravity dual 
solution in string frame is given in \cite{tdual,mr,hi}. 
Its solution is bound state solution of 
$D$3 and $D$1 branes and is given by
\ba
&&ds^2_s = f^{-{1 \over 2}}[ -dx^2_0 + dx^2_1 +h(dx^2_2 + dx^3_3)] 
+ f^{{1 \over 2}} ( dr^2 + r^2 d\Omega^2_5), \no
&&f = 1+ { {\alpha^\prime}^2 R^4 \over r^4},\;\;\; 
h^{-1} = \sin^2\theta f^{-1} + \cos^2\theta, \no
&&B_{23} = {\sin\theta \over \cos\theta} f^{-1} h, \no
&&e^{2\phi} = g^2 h,\no
&&F_{01r} = { 1 \over g} \sin\theta \partial_r f^{-1},\;\;\; 
F_{0123r} ={1\over g} \cos\theta 
h \partial_r f^{-1}.
\ea
The above solution is asymptotically flat for $r\to \infty$ and they 
have a horizon at $r=0$. In region near $r=0$ the solution has a form 
$AdS_5 \times S^5$. In order to obtain non-commutative field theory  
we should take the $B$ field to infinity.  
In the decoupling limit $ \alpha' \to 0$ with finite fixed variables
\be
 u={r \over \alpha^\prime R^2},\;\; 
 \tilde{b}=   \alpha^\prime \tan\theta,\;\;   
 \tilde{x}_{2,3}= { \tilde{b} \over \alpha^\prime} x_{2,3}, \;\;
\hat{g}= { \tilde{b}\over \alpha^\prime  }g,
\ee
and the metric becomes
\ba
ds^2 &=& \alpha^\prime R^2 \left[ u^2 \left \{ 
- d{x}^2_0+ d{x}^2_1 + \hat{h} ( d\tilde{x}^2_2 
+ d\tilde{x}^2_3 ) \right\}
+ \frac{du^2}{u^2} + d\Omega^2_5 \right], \no
\hat{h}&=&{1 \over {a+a^4 u^4}},\;\;\; a^2 =\tilde{b}R^2, \no
\tilde{B}_{23}&=&B_{\infty}{a^4 u^4 \over {1+ a^4 u^4 }},\;\;\;
B_{\infty}={\alpha^\prime \over \tilde{b} }=\alpha^\prime
{R^2 \over a^2}, \no
e^{2\phi}&=&\hat{g}^2\hat{h}, \no
A_{01}&=&\alpha^\prime {\tilde{b} \over \hat{g}}u^4 R^4, \no
\tilde{F}_{0123u}&=&{\alpha^\prime}^2 {\hat{h} \over \hat{g}}
\partial_u(u^4 R^4),
\ea 
where $\hat{g}$ is the value of the string coupling in the IR and 
$R^4 = 4\pi \hat{g} N$.
This is the gravity dual solution to NCSYM. For small $u$ which
corresponds to IR regime of gauge theory the metric reduces
to ordinary $AdS_5 \times S^5$.
This is consistent with the expectation that the noncommutative 
Yang-Mills reduces to ordinary Yang-Mills theory at long distances(IR). 
The solution starts deviating from the AdS space at 
$u \sim {1 \over a}$, i.e. at a distance scale  $ \sim R\sqrt{\tilde{b}}$. 
For large $R^4$ where supergravity limit is valid, this is greater than 
the naively expected distance scale of $L \sim \sqrt{\tilde{b}}$. 
 So the effects 
of non-commutativity is  visible at longer distances than naively expected. The
metric has a boundary at $u=\infty$. As we approach to the boundary,
the physical scale of the 2,3 direction shrinks. In this region
it seems that only $D$1 brane is relevent. The non-commutative nature arise from 
the fact that the position of D1 in D3 is not fixed but widely fluctuating.

We now discuss the symmetry property of the metric.
 In the absence of $B$ field, the metric is that of the well known $AdS_5 \times 
 S^5$.
\be
ds^2 = \alpha^\prime R^2 \left[ u^2 (-dx^2_0 + dx^2_1 + dx^2_2 
+ dx^2_3 + dx^2_3 )+{du^2 \over u^2}+ d\Omega^2_5  \right]
\ee
This metric has a rescaling symmetry at the boundary
\be
x^\mu \to \lambda x^\mu,\;(\mu =0,1,2,3)\;\; and \;\;u \to {u \over \lambda}
\ee
This symmetry is associated with the $UV/IR $: 
large   in $x$ corresponds to the small  in $u$. 
In the presence of $B$ field, however, the metric near the boundary has the form
\be
ds^2 = \alpha^\prime R^2 \left[ u^2 (-dx^2_0 + dx^2_1) + {1 \over u^2}
( dx^2_2 + dx^2_3)+{du^2 \over u^2}+ d\Omega^2_5 \right].
\ee
At the boundary the noncommutative directions shrink
and the metric effectively becomes that of $AdS_3$. 
This has  the scaling symmetry at the boundary 
\be
x^{0,1} \to {1 \over \lambda} x^{0,1},\;\;  x^{2,3} \to \lambda 
x^{2,3},\;\; and \;\;u\to \lambda u,
\ee
which is slightly different from that of the zero $B$ field case.
Therefore the co-ordinate distance $L$ along the non-commutative direction
near the boundary $u=U$ corresponds to the physical
length  ${x^{2,3}/{aU}}$\cite{dk}. 

\section{Wilson loops in various cases}
In this section we consider temporal Wilson loop at finite
temperature in non-commutative gauge theory. The gravity dual 
 is a non-extremal blackhole background with
 $B$ dependence  \cite{mr}.  
The metric is given by
\be
ds^2 = \alpha^\prime R^2 \left[ u^2 \left\{- (1-\frac{u^4_h}{u^4})
dx^2_0+ dx^2_1 + \hat{h} ( dx^2_2 + dx^2_3 ) \right\}
+ \frac{du^2}{u^2(1-\frac{u^4_h}{u^4})} + d\Omega^2_5 \right].
\label{1}
\ee
 Here tildes are omitted for convenience.
String theory on this background should provide a dual description
of non-commutative Yang Mill theory at finite temperature.
For small $u$, the metric is reduced to that of the $AdS$-Schwarzschild black hole.
Let $u_0$ be the smallest possible value of $u$ on the Wilson loop in the bulk.  
This gravity dual solution can be trusted when the following
conditions are satisfied.
\begin{itemize}
\item{small string coupling:} 
\be
e^{\phi} ={ \hat{g} \over 
\sqrt{1+a^4 u_0^4}} \ll 1,
\ee
\end{itemize}
\begin{itemize}
\item{small curvature :}
\be 
 \hat{g}^2_{YM}N = \hat{g}N \gg 1.
\ee
\end{itemize}     

We know from the above gravity solution, that
$au_0 \sim 1$ is a transition region from  $AdS_5$ blackhole region to dimensionally 
reduced  $AdS_3$ region.  
 Let $u_0$ be the minimal value available to the string 
 configuration.  The noncommutative effect
is relevant to $au_0 \gg 1$(UV) region.
Since 
it is expected to get Coulomb's law for $au_0 \ll 1$(IR) where the 
noncommutaive effect is invisible, it is expected to get something else for 
the quark-antiquark potential, like an area law.  We will show 
indeed this is so by calculating the Wilson loops in various cases and also we will find 
necessary condition for this to happen. 
There is agreements that IR behaviour is Coulombic \cite{mr,aos,dk}, 
while there are different opinions for the UV behavior. Therefore
our interest will 
be for  $au_0 \gg 1$(UV) where the noncommutative effect is 
manifest in the metric behavior ($a\sim {\tilde b}$).

%%%%%%%%%%%%%%%%%%%%%%%%%%%%%%%%%%%%%%%%%%%%%%%%%%%%%%%%%%%%%%%%%
\subsection{Wilson loops in non-commuative gauge 
theory at finite temperature }

\subsubsection{Temporal loop}
%%%%%%%%%%%%%%%%%%%%%%%%%%%%%%%%%%%%%%%%%%%%%%%%%%%%%%%%%%%%%%%%%

Now consider classical string world-sheet action, Nambu-Goto action,
which is given by
\be
S = {1\over {2\pi\alpha^\prime}} \int d\sigma d\tau
\sqrt{det(h_{\alpha\beta})},
\ee
where $h_{\alpha\beta} =G_{MN} \partial_\alpha X^M 
\partial_\beta X^N$.
We choose static gauge as $ \tau=x_0,  
\sigma=x_2 ,u= u(\sigma)$.
Then on the background of (\ref{1}), we have  
\ba
&&h_{\tau\tau} = \alpha^\prime R^2 u^2 ( 1-\frac{u^4_h}{u^4} ), \no
&&h_{\tau\sigma} = h_{\sigma\tau} =0, \no
&&h_{\sigma\sigma} = \alpha^\prime R^2 u^2 \hat{h} + \alpha^\prime
u^{-2} (1-\frac{u^4_h}{u^4})^{-1} (\partial_\sigma u)^2.
\ea
So the Nambu-Goto action can be written as
\be
S= \frac{R^2}{2\pi} T \int dx_2 \sqrt{ \hat{h} ( u^4-
u^4_h) + (\partial_{x_2} u)^2 }. \label{action1}
\ee
The action does not explicitly depend on $x_2$ so it gives
\be
\frac{\hat{h} (u^4-u^4_h) }{\sqrt{
\hat{h} ( u^4-u^4_h  ) + (\partial_{x_2}u)^2 } } = c.
\ee
At $u=u_0$, where it is the closest point to the horizon,
$\partial_{x_2} u =0$ and $\hat{h} \to \hat{h}_0$. Then we can 
determine $c$
\be
c^2= \hat{h}_0 ( u^4_0 - u^4_h).
\ee
This allows us to write $x_2$ as a function of $u$
\be
x_2= \sqrt{{u^4_0-u^4_h}\over {1+a^4u^4_h}}\int du { {1+a^4 u^4}\over 
{\sqrt{(u^4-u^4_0)(u^4-u^4_h)} } }.\label{213}
\ee
For large $u$,
\be
x_2 \sim a^4 \sqrt{ \frac{u^4_0 - u^4_h}{1+ a^4 u^4_h} }u = ku,
\ee
where $k$ can be interpreted as a slope for large $u$. 
This implies that we cannot fix the position of the string at 
infinity since $x_2$ grows linearly with $u$. See figure 1.
\begin{figure}[hbt]
\centerline{\epsfig{file=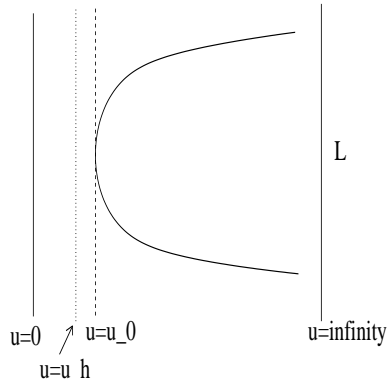,width=5cm,height=5cm}}
\caption{\small $Q{\bar Q}$ separation $L$ grows indefinitely since
$x_3=k u$ as we approaches to the boundary.} 
\end{figure}
This dependence of 
$u$ is associated to non-locality\cite{bs}
of noncommutative theory. 
In the IR region  we get the Coulombic potential by similar calculation 
of  \cite{aos}. However, we will soon see that in the UV region ($au_0\gg1$) we get different
one.  

L and $u_0$ is related by
\be
{L\over 2}=x_2(u\to \infty)= \sqrt{ {{1- {u^4_h \over u^4_0}}
\over {1+ a^4 u^4_h}} } {1\over u_0} \int^\infty_1 dy 
{ {1+a^4 u^4_0 y^4} \over {\sqrt{(y^4-1) 
(y^4-{u^4_h \over u^4_0} )} }},
\ee
where $y={u \over u_0}$. The energy of this
configuration  is given by
\be
E= {R^2 \over \pi}\sqrt{{{1+ a^4 u^4_0}\over {1+ a^4 u^4_h}}}u_0
\int^\infty_1 dy \sqrt{{{y^4-{u^4_h\over u^4_0}}\over {y^4-1}}}.
\ee

For $u_ha \le u_0a \ll 1$,
it is known that we get the screened Coulomb potential \cite{fiwil,wloop}.  \\
We are currently interested in the UV regime ($au_0\gg1$) where the noncommutative effect
is crucial. 
When we consider $au_0 \gg 1$  case (equivalently $u_h/u_0 \to 0$), $L$ and $E$ are
respectively given by 
\be
{L \over 2} \sim {1 \over u_0} \left[ a^4 u^4_0 \int^{\infty}_1
 dy {y^2  \over {\sqrt{y^4 -1}}} \right],\;\;{\rm and} \;\;
E \sim {R^2 \over \pi} a^2 u^2_0\cdot u_0 \int^{\infty}_1 dy 
 {y^2  \over \sqrt{y^4 -1}}.
\ee
Then the relation between $E$ and $L$ can be read. 
\be
E={R^2 \over{\pi a^2}}L  \label{EL}
\ee
Both separation length and energy are infinite and there is no canonical way to renormalize the 
separation length. If all what we want is the $E-L$ relation, the 
only reasonable way to regularize both  quantity is to put a 
cutoff in $u$. The question is whether we can give a physical 
interpretation to this. We will try this at section 4.
For a moment, we just calculate the the Wilson loops for various 
other situations. 

\subsection{Spatial loop}
Spatial Wilson loops in high temperature are interesting since they can be considered 
as ordinary Wilson loops of Euclidean field theory in one less 
dimension. 
\subsubsection{In three dimension} %%%%%%%%%%%%%%%%%%%%%%%%%%%%%%

Let us start with the Euclidean non-extremal D3-brane metric with
$B_{23}$ fields. The metric has the form as before
\be
ds^2 = \alpha^\prime R^2 \left[ u^2 \left\{ (1- {u^4_h \over u^4} 
)dt^2_E
+dx^2_1 + \hat{h}(dx^2_2 + dx^2_3) \right\} +
{du^2 \over {u^2 (1-{u^4_h\over u^4})}} + d\Omega^2_5 \right].
\ee
 Compactify
this four dimensional theory on ${\bf{S}}^1$  of $t_E$ by periodically
identify its period with the inverse Hawking temperature  
proportional to $u_h$ and take high temperature limit so
that the radius becomes zero. The resulting effective theory is
then interpreted as Euclideanized three dimensional noncommutative 
field theory.
The circle compactification breaks both supersymmetry and conformal symmetry.
After compactifying $t_E$, the resulting three dimensional theory
is described by coordinates $x_1, x_2$, and $x_3$. Among them,
for instance, $x_1$ can be considered as  a Euclideanized time and $x_2, x_3$ as  
spatial coordinates. For the spatial loop, the string configuration we use is 
$\tau=x_1$ ,$\sigma=x_2$ and $u=u(\sigma)$.
Then  the Nambu-Goto action become
\be
S= {T R^2\over{2\pi}} \int dx_2 \sqrt{
u^4 \hat{h} + (1- {u^4_h\over u^4} )^{-1}(\partial_{x_2} u)^2 },
\label{action2}
\ee
where $\hat{h} = {1\over { 1 + a^4 u^4}}$.  
Then it is easy to show that the separation length  
\be
L= {2\over u_0} \int^{\infty}_1 dy 
{ {1+ a^4 u^4_0 y^4 }\over {\sqrt{(y^4-1)
(y^4- {u^4_h\over u^4_0}) }} }.
\ee
 and  energy 
\be
E={ R^2 u_0 \sqrt{1+a^4 u^4_0} \over{2\pi} } \int^{\infty}_1
dy { y^4\over {\sqrt{(y^4-1)(y^4- {u^4_h\over u^4_0})}}}.
\ee
>From these, the relation between $E$ and $L$ when the noncommutative
parameter $au_0$ is  
\be
E \sim {R^2 \over {2\pi a^2}} L.
\ee
In case of $B=0$, 
 the spatial loop is
gives an area law, while the temporal loops are not. 
Here ($B \ne 0$), both give the area law when the noncommutative parameter is 
large.

%%%%%%%%%%%%%%%%%%%%%%%%%%%%%%%%%%%%%%%%%%%%%%%%%%%%%%%%%%%%%%%%%%%%%%%%
\subsubsection{ Four dimensional case}
%%%%%%%%%%%%%%%%%%%%%%%%%%%%%%%%%%%%%%%%%%%%%%%%%%%%%%%%%%%%%%%%%%%%%%%%

Here we 
start  with non-extremal $D$4-brane.  The metric
with $B$ fields\cite{mr} is given by
\be
ds^2 = f^{-{1\over 2}} [ dx^2_0 + dx^2_1 + h(dx^2_2 + dx^2_3) 
+ dx^2_4] + f^{{1\over 2}} (dr^2 + r^2 d\Omega^2_4),
\ee
where
\ba
&&f= 1+ { {\alpha^{\prime}}^{{3\over 2}}R^3 \over r^3},
\;\; h^{-1} =f^{-1}  \sin^2\theta + \cos^2\theta,\no
&&B_{23} =f^{-1}h \tan\theta,\no
&&e^{2\phi}=g^2 f^{{-1\over 2}} h.
\ea
The decoupling limit exists and is given by
\ba
&&r = \alpha^\prime R^2 u,\;\; \alpha^\prime \to 0,\no
&&\tan\theta = {\tilde{b}\over \alpha^\prime},
\;\;x_i = {\alpha^\prime \over {\tilde{b}} }\tilde{x}_i, 
\;\;a^3 =\tilde{b}^2  {\alpha^\prime}^{-1\over 2} R^3, \no
&&\tilde{B}_{23} ={\alpha^\prime \over \tilde{b} }{ {a^3 u^3}
\over {1+a^3 u^3} }.
\ea
For non-extremal case the metric is simply written as
\ba
ds^2 = &&\alpha^\prime R^2 \left[ ({\alpha^\prime}^{{1\over 4}}
R^{{1\over 2}} u^{{1\over 2}})^{-1}u^2 \left\{
(1- {u^3_h \over u^3}) dt^2 + dx^2_1 + \hat{h} (dx^2_2 + dx^2_3)
+ dx^2_4 \right\} \right .\no
&&\left. +( {\alpha^\prime}^{{1\over 4}}
R^{{1\over 2}} u^{{1\over 2}}) ( {du^2\over {u^2(1-{u^3_h \over u^3})} } 
+ d\Omega^2_4) \right],
\ea
where $\hat{h} = {1\over {1+ a^3 u^3}}$ and tildes are ommited for convenience.
For static string configuration $\tau=x_1 , \sigma=x_2, u=u(\sigma)$, 
the Nambu-Goto action can be written as
\be
S = {T R^2 \over {2\pi}} \int dx_2 \sqrt{ {\alpha^\prime}^{
{-1\over 2}} R^{-1}u^3 \hat{h} + (1- {u^3_h \over u^3} )^{-1} 
(\partial_{x_2} u)^2}.
\ee

The first integral is 
\be
{ { {\alpha^\prime}^{{{1\over 2}}} R^{-1} u^3 \hat{h} } \over
{ \sqrt { {\alpha^\prime}^{{-1\over 2}} R^{-1} u^3 \hat{h}
+(1- {u^3_h\over u^3})^{-1} (\partial_{x_2} u)^2 }}} =c,
\ee
where $ c^2 ={\alpha^\prime}^{{-1/2}} R^{-1} u^3_0 \hat{h}_0$.
After some calculation, the separation length
\be
L = { 2 u^{{-1\over 2}}_0 \over { \sqrt{ {\alpha^\prime}^{
-1\over 2}R^{-1}}} }\int^{\infty}_1 dy { {1 + a^3 u^3_0 y^3 } \over
{ \sqrt{ ( y^3-1)( y^3 - {u^3_h \over u^3_0}) }}},
\ee
and the energy  
\be
E= { R^2 u_0 \sqrt{1+ a^3 u^3_0} \over \pi } \int^\infty_1 dy
{ y^3 \over { \sqrt{ (y^3 -1)(y^3 - {u^3_h\over u^3_0}) }} }.
\ee
For UV-regime $au_0 \gg 1$, we again get an area law.
\be
E = \left ( R^3 \over { {\alpha^{\prime^{1\over 2}} \pi^2 a^3 }} 
\right)^{1\over 2} L.
\ee 
%%%%%%%%%%%%%%%%%%%%%%%%%%%%%%%%%%%%%%%%%%%%%%%%%%%%%%%%%%%%%%%%
\subsection{Wilson loop in $D$-instanton background}
%%%%%%%%%%%%%%%%%%%%%%%%%%%%%%%%%%%%%%%%%%%%%%%%%%%%%%%%%%%%%%%%

In this section we consider one more example:  quark-antiquark potential in  
 $D$-instanton background with constant $B$ field. 
The gravity dual solution for $B=0$ was considered 
in \cite{hliaat}. It is 
easy to turn on $B$ field for this background. First rotate
the $D$-instanton background and then $T$-dualize it.
Then the resulting metric\cite{rs2} is given by
\be
ds^2 = H^{{1\over 2}} \left[ f^{{-1\over 2}} \left\{
dt^2 + dx^2_1 + h(dx^2_2 + dx^2_3) \right\} + f^{{1\over 2}}
(dr^2 + r^2 d\Omega^2_5) \right],
\ee
where
\ba
 &&f= 1+  {{\alpha^\prime}^2 \over r^4 },\;\;
H = 1+ {{ q {\alpha^\prime}^4 R^4}\over r^4 }, \no
&& e^{2\phi} = g^2 h H,\;\; B_{23} = f^{-1 }h H\tan\theta, \no
&&h= { 1\over {H f^{-1} \sin^2\theta + \cos^3\theta} }.
\ea
This $D$-instanton is smeared over $D3$ brane worldvolume. 
This solution is $T$-dual to $D$4+$D$0 or $D$5+$D$1(with $B$ field) 
brane configuration. In the decoupling limit
\ba
&& r= \alpha^\prime R^2 u,\;\; \alpha^\prime \to 0 ,\no
&&\tan\theta = {\tilde{b} \over \alpha^\prime} ,\;\;
f \to ({\alpha^\prime}^2 R^4 u^4)^{-1},\;\;H \to 1+ {q \over {R^4 u^4}},
\;\;a^2 =\tilde{b} R^2, \no
&&h \to  { {\tilde{b}}^2 \over { {\alpha^\prime}^2
( 1+ H a^4 u^4)} },\;\;
B_{23} \to { H {a^4 u^4 \over {(1+ H a^4 u^4)}} }, \no
&&x_{0,1} =\tilde{x}_{0,1},\;\; x_{2,3} = {\alpha^\prime\over 
{\tilde{b}}}\tilde{x}_{2,3}.
\ea
So the metric becomes
\be
ds^2 = \alpha^\prime R^2 ( 1+ { q\over {R^4 u^4}})^{{1\over 2}}
\left[ u^2 \left\{d\tilde{x}^2_0 + d\tilde{x}^2_1 + \hat{h}
( d\tilde{x}^2_2 +d\tilde{x}^2_3) \right\} +
({du^2\over u^2} + d\Omega^2_5) \right], \label{ins}
\ee
where $\hat{h} = {1\over {1+ H a^4u^4}} $ and $H= 1+ {q\over {R^4 u^4}}$.
The metric becomes flat ${\bf R^{10}}$ as $u\to 0$ 
while it deviates from flat as $u$ become large. 
When $B$ field is zero, this metric solution describes
wormhole solution which connects flat space(${\bf R}^{10}$) in $u \to
0$ with $AdS^5 \times S_5$ in $u \to \infty$.

For the static string configuration, $\tau=x_0 , \sigma=x_2$,
and $u=u(\sigma)$,
the Nambu-Goto action can be written as
\be
S= { R^2 T \over {2\pi}} \int dx_2 \sqrt{ H
 \{ u^4 \hat{h} +(\partial_{x_2} u)^2 \} },
\ee
which gives the following first integral.
\be
{ \sqrt{H} u^4 \hat{h} \over \sqrt{ u^4\hat{h} 
+(\partial_{x_2} u)^2 } } =c,
\ee
where $c^2 = H_0 u^4_0 \hat{h}_0$.
$L$ and $E$ are given by
\ba
{ L \over 2}  &=& {\sqrt{(1+ { q\over {R^4 u^4_0}})
}\over u_0} \int^\infty_1dy {  (1+ a^4 y^4 u^4_0 +{ qa^4 \over R^4}) 
\over{ y^2 \sqrt{ y^4-1} } }, \no
E&=& {R^2 \over {\pi}} \sqrt{ 1+ a^4 u^4_0 + {qa^4 \over R^4} } u_0
\int^\infty_1 dy {(y^2  + { q \over {R^4 y^2 u^4_0} }) \over
\sqrt{y^4-1}} .
\ea
Here our  interest
lies in the region where  $au_0 \gg 1$ and $u_0 \gg u_h $,  
\be
L \sim a^4 u^3_0 \int^{\infty}_1
dy {y^2 \over {\sqrt{ y^4-1}}},\;\;and\;\;
E \sim { a^2 R^2 u^3_0 \over \pi} \int^{\infty}_1
dy { y^2 \over {\sqrt{y^4-1}}},
\ee 
which after elimination of $u_0$ gives 
\be
E= {R^2 \over {\pi a^2}} L  \label{diel1}. 
\ee
One should notice that in UV region, $u>u_0>1/a$, the D-instanton 
effect is negligible and the non-commutativity effect is dominant. 
The fact that the D-instanton charge is proportional to $\alpha'^4$ 
is crucial to neglect the D-instanton effect in the UV region. 

For $B=0$, it is known  \cite{hliaat} that the large 
$L$ correspond to ($u_0 \to 0$) and it leads to an area law. To 
compare this and above case let's look at some details.
In this case $L$ and $E$ is given by
\ba
{L \over 2} &=& \sqrt{ 1+ {q \over {R u^4_0}}} {1 \over u_0}
\int^\infty_1 dy {1\over { y^2 \sqrt{y^4-1}}}=
{ 2\sqrt{2}\pi^{3\over2}  \over \Gamma({1\over4})^2 }  
\sqrt{ 1 + {q \over {R^4 u^4_0}}} {1\over u_0}, \no
E &=& {R^2 \over \pi} \left[ \int^\infty_1 dy \left( { y^2 \over 
{\sqrt{y^4-1}}} -1 \right)-1 \right] + {R^2 \over \pi} {1 \over u^3_0}
\int^\infty_1 dy {1 \over {y^2 \sqrt{y^4-1}}} \no
&=& -{ 2 R^2 u_0 \sqrt{\pi} 
\over \Gamma({1\over 4})^2} + {\sqrt{2} R^2 \sqrt{\pi} \over 
{\Gamma({q\over 4})^2 u^3_0}}.  
\ea
As $u_0 \to 0$, $L \to \infty$, we have
\be
E ={R^4 \over {2\pi \sqrt{q}} } L \label{diel2}.
\ee
Although  \ref{diel1} and  \ref{diel2} gives similar results the 
details are very different. $B=0$ case is the IR result ($u_0 \ll 1$) and 
is a consequence of dynamics, while $B \ne 0$ case is the UV result 
and  kinematical. 

\subsection{General $D_p$ brane cases}
In order to study  more general case, we look 
at the Wilson loop in higher dimensional non-commutative Yang-Mill 
theories. The dual gravity metric is given by \cite{mr}
\be
ds^2 = H_p^{-1/2} [ h_0 (-dx^2_0 + dx^2_1) +h_1(dx^2_2 + dx^2_3)  + \cdots] +H_p^{1/2} [du^2 +u^2 d\Omega^2_5 ]
\ee
with $H_p \sim 1/u^{7-p}$ and $h_i=1/(a_iu)^4$.
We again, do not need to consider the black holes since we are interested in large $au_0 \gg 1$ case.
The general string action is
\be
S \sim T \int dx \sqrt{h_0( H^{-1}h +(\partial_x u)^2 ) }. \label{oa}
\ee
Here we assumed that the string is along $x$ direction which is one of the  non-commutative planes.
We immediately see that as $u\to \infty$, $ H^{-1}h \to constant$, resulting in the linear potential
if and only if $h_0 =constant$, namely of no $B_{01}$ field is applied.  
This can be easily verified if we observe that when $h_0$ is constant
 the action (\ref{oa}) leads to the first integral
 \be
 (\partial_x u)^2 + f(1-f/f_0)=0
 \ee
 where $ f(u)=H^{-1}(u)h(u) $ and $f_0$ is the value of $f$ at $u=u_0$. 
 The qualitative  behavior of the solution can be read off from the particle 
 moving with zero energy under the potential   $f(1-f/f_0)$. See 
 figure 2. 
 The effect of D3 brane charge  ($F_5$ flux $H(u)$) is to pull out the particle to the boundary of AdS,
 while that of NS-NS charge ($B_{23}$ field $h(u)$) is to  pull in the particle into 
 the horizon. In terms of boundary variable, the former expands the $x^\mu$ along the brane
  directions, while the latter shrink the $x^{2},x^3$ plane.  {\it The essence of 
the phenomena is the exact cancellation of two effect in the asymptotic region.}
Since both $H$ and $h$ are based on the harmonic power $u^{-(7-p)}$ of the transverse dimensions, 
this is unavoidable in the region where those terms are dominant.  
The net effect is such that the particle has a constant speed, or 
the Wilson loop has a constant slope. 

\begin{figure}[pot]
\centerline{\epsfig{file=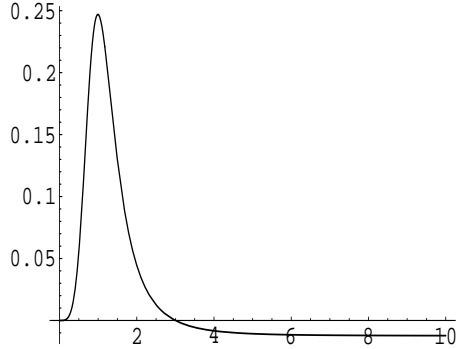,width=6cm,height=5cm}}
\caption{The minimal surface problem is reduced to the motion of a classical particle 
moving with 0 total energy in a potential $V= f(1-f/f_0)$. 
The potential  peak is near the $1/a$. In the plot, we set $a=1$, $u_0=3$.} 
\end{figure}

\section{D1 dominance and Dimensional reduction}
Now we want to figure out the origin of the linear 
behavior of the potential in more physical terms for the interested D3 case. 
Consider the metric for following class: 
\be
ds^2 = \alpha^\prime R^2  \left[ u^2 h_0 (-dx^2_0 + dx^2_1) +u^2 h_1(dx^2_2 
+ dx^2_3 + dx^2_3 )+{du^2 \over u^2}+ d\Omega^2_5  \right]
\ee
where $h_i=1/(1+(a_iu)^4)$.  For large $au_0 \gg 1$ and  $u_0 \gg u_h$, we 
do not need to care about the black hole effect. The 
non-commutativity effect ($B$ field effect) is dominant. 
The string action is 
\be
S \sim T \int dx_2 \sqrt{h_0( u^4h_1 +(\partial_x 
u)^2 ) }. \label{oa1}
\ee
For the case we considered, $h_0 =1$ and $h_1 \sim 1/u^4$, so that 
the action becomes
\be
S\sim T/a \int \sqrt{1+(\frac{du}{dx})^2} 
, \label{linear}\ee
after the  scaling $u \to u/a$ and  $x\to ax$.  
Therefore the energy is proportional to the line element of flat 
space. The shortest length is for the straight line. 
Therefore since the line is not orthogonal to the boundary, 
the energy has to be proportional to the  linear 
length in the x-direction. 
What happen  if we turn 
on $B_{01}$ also? In this case non-trivial $h_0$ arise \cite{mr}. 
The metric for large $u$ region is that of $AdS_5 \times S^5$  \cite{mr};
\be
ds^2 = \alpha^\prime R^2  \left[ { -dx^2_0 + dx^2_1 + dx^2_2 
+ dx^2_3 + du^2 \over u^2}+ d\Omega^2_5  \right]
\ee
which is the  metric of $AdS_5 \times S^5$ ( but in Poincare co-ordinate). 
So it is similar to the near horizon  geometry of distributed D-instanton over 
the D3 \cite{mr,hliaat,psin}. 
The boundary $u=\infty$ is now at the AdS-horizon.
In fact one can show that  the critical path with 
the given boundary condition does not exist. 
From the calculational point of view, what is crucial for the area law
is the absence of $h$ factor in the  
$g_{tt}$. However this means that
$g_{11}$ is also free of  the $h$ factor. Therefore we have to have one 
spatial direction along which $B$ field is not applied to get the 
area law. According to the super gravity solution (2.1),  there is a D1 branes along $x^1$ 
direction. Furthermore, near the boundary, 
the existence of D3 brane is suppressed by 
$h \sim 1/u^4$ factor relative to the D1 branes. 
\be
F_{01r} = { 1 \over g} \sin\theta \partial_r f^{-1},\;\;\; 
F_{0123r} ={1\over g} \cos\theta 
h \partial_r f^{-1}.
\ee
In near horizon limit, 
\ba
F_{01u}&=&\alpha^\prime {\tilde{b} \over \hat{g}}\partial_u (u^4 R^4), \no
\tilde{F}_{0123u}&=& \hat{h}\cdot {{\alpha^\prime}^2  \over \hat{g}}
\partial_u(u^4 R^4)
\ea
Another manifestation of the D1 dominance\cite{roy,cai} near the boundary is the metric itself. 
The near horizon limit of the metric shows that $g_{22}, g_{33}$ 
is suppressed by the same factor $h \sim 1/u^4$ compared to the 
$g_{11}$. In fact this suppression of the non-commutative 
direction is the motivation to begin this work. 
The non-commutativity in $x^2,x^3$ can be interpreted as the 
fluctuation of the the location of the D1 brane along those 
directions. In fact this is origin of the fluctuation of the end 
point of the Wilson line noted in \cite{mr}. 

One may further understand the behaviour of Wilson loop by considering the 
open string as dipole \cite{jabbari,susskind} and taking the analogy to 
the charged particle in magnetic field.  
In case of charged particle,  
when $F_{23}$is applied,  the particle stay in the lowest 
Landau level and only transverse $x^1$ direction 
is available for the free motion. This is so called 'dimensional 
reduction' due to the magnetic field. The particle moves in the 
effective 1+1 dimension whose  kinematic effect gives Coulomb's 
law of linear potential. 
This is closely parallel to the fact in metric: $g_{22}$ and 
$g_{22}$ is highly suppressed relative to $g_{tt}$ and $g_{11}$.

\section{Summary and Discussion}
In this paper, we  study  the UV behaviour of the
Wilson loop in the non-commutative gauge theory.   The Wilson loop 
calculation in AdS/CFT is reduced to the particle dynamics in a 
potential defined by the  D3 brane charge and NS-NS $B$ 
field. In spite of the  the lack of the 
gauge invariance of the Wilson loops in non-commutative gauge theory,
a physically meaningful aspect of Wilson loop comes out. 
After calculating various cases, we observed that the area law in the UV region is 
universal if no $B_{01}$ is applied and it is 
 consequence of balance of two competing tendency: 
 the effect of $F_5$ flux ($H(u)$) is to pull out the particle to the boundary of AdS,
 while that of $B_{23}$ ($h(u)$) is to  pull in the particle into 
 the horizon. 

In case of D3 brane, the effect has striking similarity with so 
called dimensional reduction and 'Magnetic Catalysis', where strong magnetic field project 
the electron states to its lowest Landau level so that the charged 
particle has reduced degrees of freedom: it is effectively 1+1 dimensional 
system\cite{miranski,hsin}.
If the magnetic field $F_{23}$ is turned on, the $x^2,x^3$ plane is
effectively confining the electron motion   and the system undergoes 
dimensional reduction, which in turn causes chiral symmetry breaking of a massless
fermion system. Apparently, 
the similarity between the charged particle and open string in strong magnetic 
field is not complete, since the string is dipole rather 
than a charge. If the string aligned along the $x^1$ direction 
transverse to the non-commutative plane, it does not see the 
dimensional reduction at all. However, the Wilson line we discussed is with zero velocity
and the particle with zero 
velocity does not feel any magnetic field nor the dimensional 
reduction, either. So, the parallelism is stronger than expected. 
So, it would be interesting to study whether magnetic catalysis phenomena exist in the 
3+1 dimensional non-commutative field theory.
 
\vskip 1cm 
\noindent{\bf Acknowledgements}\\
This work was supported by Korea Research Foundation Grant
(KRF-2000-015-DP0081) and  by Brain Korea 21 Project in 2001.

\end{document}